\documentclass[aip,pop,preprint,graphicx,amsmath,amssymb]{revtex4-1} 
\usepackage{graphicx} 
\usepackage{rotating}

\usepackage{color}

\newcommand{\helios}{HELIOS-CR}
\newcommand{\propaceos}{PROPACEOS}
\newcommand{\raven}{RAVEN}

\begin{document}


\title{One-dimensional radiation-hydrodynamic simulations of imploding spherical plasma
liners with detailed equation-of-state modeling} 

\author{J. S. Davis} 
\altaffiliation{Now at Applied Physics, University of Michigan, Ann Arbor, MI 48109}
\author{S. C. Hsu} \email[Email: ]{scotthsu@lanl.gov} 
\affiliation{Physics Division, Los Alamos National Laboratory, Los Alamos, NM 87545} 

\author{I. E. Golovkin}
\author{J. J. MacFarlane}
\affiliation{Prism Computational Sciences, Inc., Madison, WI 53711}

\author{J. T. Cassibry}
\affiliation{Propulsion Research Center, University of Alabama in Huntsville,
Huntsville, AL 35899}


\date{\today}

\begin{abstract} 

This work extends the one-dimensional radiation-hydrodynamic imploding
spherical argon plasma liner simulations of T. J. Awe {\em et al.}
[Phys.\ Plasmas~{\bf 18}, 072705 (2011)] by using a detailed
tabular equation-of-state (EOS) model, whereas Awe {\em et al.}\ used
a polytropic EOS model.
Results using the tabular EOS model give lower stagnation pressures by a
factor of 3.9--8.6 and lower peak ion temperatures compared to
the polytropic EOS results.
Both local thermodynamic equilibrium (LTE) and non-LTE EOS models were used
in this work, giving
similar results on stagnation pressure.
The lower stagnation pressures using a tabular EOS model
are attributed to a reduction in the liner's ability
to compress arising from the energy sink introduced by ionization and
electron excitation, which
are not accounted for in a polytropic EOS model.
Variation of the plasma liner
species for the same initial liner geometry, mass density, and velocity
was also explored using the
LTE tabular EOS model, showing that the highest stagnation pressure
is achieved with the highest atomic mass species for the constraints imposed.

\end{abstract}



\maketitle




\section{Introduction} 
\label{introduction}
The Plasma Liner Experiment (PLX) at Los Alamos
National Laboratory was designed to form and study
spherically imploding plasma liners via the merging of thirty supersonic
plasma jets.\cite{hsu12}  Such imploding plasma liners could potentially
allow repetitive generation of cm-, $\mu$s-, and Mbar-scale plasmas for
fundamental high energy density plasma physics studies, or (if scaled up
in energy) be used as 
a standoff compression driver\cite{thio99,thio01,hsu12} for magneto-inertial fusion
(MIF).\cite{lindemuth83,kirkpatrick95,lindemuth09}
Recent one-dimensional (1D) radiation-hydrodynamic simulations
of targetless imploding spherical argon plasma liners by Awe {\em et al.}\cite{awe11}
used a polytropic
equation-of-state (EOS) model, providing insight into the 1D physical
evolution of imploding spherical plasma liners, and showing that 
PLX-relevant plasma liner
kinetic energies of hundreds of kJ could result in liner
stagnation pressures on the order of 1~Mbar sustained for order 1~$\mu$s.
The results of Awe {\em et al.}\cite{awe11}
indicated high enough plasma temperatures (10's of eV) for
ionization effects, which are not included in a polytropic EOS model,
to have an appreciable impact on both the predicted peak pressures and
temperatures.   As some of the initial liner kinetic energy would now be used
for ionization, it is expected that the liner may 
compress less, although the expected drop in thermal energy in the imploding
plasma liner due to 
ionization and radiative loss channels are expected to help compensate.   
Note that this work and that of Awe {\em et al.}\cite{awe11}
focus on targetless, self-collapse
of the liner, motivated by the goals of the PLX project.
Thus, care must be taken in using these results to provide insight into
plasma liner implosions onto a magnetized target for MIF, which
has been explored in several
other recent publications.\cite{parks08,cassibry09,samulyak10,
santarius12}

The purpose of this work is to develop insight into how
a detailed EOS model, and particularly ionization, affect spherically
symmetric, targetless plasma liner implosions.
We used the \helios\cite{macfarlane06}
1D radiation-hydrodynamic code and the \propaceos\cite{macfarlane06}
EOS and multi-group opacity database to repeat the simulation cases
presented in Awe {\em et al.}\cite{awe11}  In particular, we examined
how the detailed EOS model affected the liner stagnation pressure $P_{stag}$
(defined in Sec.~\ref{benchmarking}),
the stagnation time $\tau_{stag}$ (defined in Sec.~\ref{benchmarking}),
the implosion trajectory and minimum
radius achieved, and scaling behavior of $P_{stag}$ 
as a function of initial plasma liner parameters.  In all cases,
we used argon as did Awe {\em et al.}\cite{awe11}  However, for a particular
initial liner geometry, mass density, and velocity (corresponding to
an initial energy of 375~kJ, relevant for PLX),
we also ran simulations using H, D-T, $^4$He, $^6$Li, $^{11}$B, Ne, Kr, and Xe,
in addition to Ar.
In this work, as in Awe {\em et al.},\cite{awe11} we set aside the important
issues of 3D effects of
the discrete merging plasma jets and 3D convergent instabilities
such as Rayleigh-Taylor and others, which are addressed elsewhere,\cite{cassibry12}
and study spherically
symmetric imploding plasma liners initiated at a jet merging radius $R_m$ at which 
plasma jets are assumed to have merged.

The remainder of the paper is organized as follows.  Sec.~\ref{approach}
describes our modeling approach including the choice of liner initial
conditions investigated, details of the codes used, and benchmarking
against the results of Awe {\em et al.}\cite{awe11} Section~\ref{results} presents
the new simulation results including
(1)~comparison of results using
local thermodynamic equilibrium (LTE) versus non-LTE EOS tables, (2)~comparison
of plasma liner
implosions using a LTE tabular EOS versus a polytropic EOS, (3)~$P_{stag}$ scaling 
with initial liner density and velocity, and (4)~different plasma species for one
liner initial condition.  Section~\ref{summary} provides a discussion and summary.

\section{Modeling approach}
\label{approach}

In this section we describe the simulation initial 
conditions, the codes used, and benchmarking of our own
polytropic EOS results against the results of Awe {\em et al.}\cite{awe11}
 
\subsection{Simulation initial conditions}

\begin{figure}[!h]
\includegraphics[width=2.5truein]{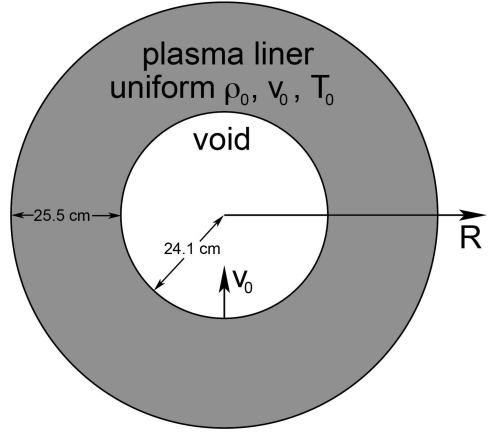}
\caption{Schematic of the plasma liner initial conditions used in all simulations.
The initial liner mass densities $\rho_0$ and velocities $v_0$ are given in 
Table~\ref{eos-table}\@.  The initial liner temperature $T_0 = T_i = T_e$ is 1.0~eV\@.}
\label{setup}
\end{figure}

The initial conditions used in this work are shown
schematically in Fig.~\ref{setup} and summarized in Table~\ref{eos-table}
(columns 2--5).
The rationale for these choices are described in
detail in Awe {\em et al.}\cite{awe11}  In short, they are intended
to span a range of initial plasma liner velocities $v_0$ and
kinetic energies $KE_0$ from PLX-relevant values (50~km/s, hundreds of kJ) at
the lower end to what
might be needed for an MIF standoff compression
driver\cite{thio99,thio01,hsu12} ($>50$~km/s, $>$~tens of MJ) at the higher end.
The initial liner ion density values $n_0$ are what can be
expected based on plasma jet densities that
have already been achieved by the plasma guns designed and built by HyperV
Technologies\cite{witherspoon11} for use on PLX\@.
In this work, we assume that discrete plasma jets have merged
at $R_m=24.1$~cm and formed a spherically symmetric imploding
plasma liner with uniform mass density $\rho_0$, thickness 25.5~cm,
temperature $T_0=T_e=T_i=1.0$~eV, and uniform inward radial velocity $v_0$.  
The effects of non-uniform initial liner profiles are not studied here but
are a potential way of optimizing imploding plasma liner performance.\cite{kagan11}
The void region (Fig.~\ref{setup}) is modeled using very low density plasma 
with mass densities of $10^{-8}$~kg/m$^3$ (tabular EOS runs) and $10^{-9}$~kg/m$^3$ 
(polytropic EOS runs), temperature of 1.0~eV, and the same species
as the liner.  The different initial void mass densities led to slightly different 
computational zone setups but did not affect the overall liner evolution.

Although this work does not focus on the
physical evolution of a spherical 
plasma liner imploding on vacuum, which was described in detail
in Awe {\em et al.},\cite{awe11}  we provide here a brief summary
of its 1D evolution:  (1)~the imploding
liner converges toward the origin, and the liner density rises as $R^{-2}$, (2)~upon
the liner reaching the origin, an outward going shock propagates
radially outward, shock-heating the incoming liner material, and (3)~the
outgoing shock reaches the outer edge of the incoming liner, at which time the 
high pressure of the post-shocked liner material can no longer be sustained, and the
entire system disassembles.

\subsection{Computational codes}
\label{codes}

This work used the \helios\cite{macfarlane06} 1D radiation-hydrodynamic code
with LTE and non-LTE EOS tables generated using the
\propaceos\cite{macfarlane06} code.  
\helios\ is a Lagrangian 
code that can use both a polytropic
EOS (with adiabatic constant $\gamma$)
or a tabular EOS\@.  
The \helios\ code has been applied previously to the modeling of inertial
confinement fusion (ICF) implosions.\cite{macfarlane05,welser-sherrill07}
\propaceos\ EOS and frequency-dependent opacities are computed using a combination
of an isolated atom model and, at high densities at which cohesive effects are
important, a quotidian-EOS-like model.\cite{more88}  At the densities of interest in
this paper, energies and pressures are computed for an isolated atom model in which
atomic level populations are computed using Boltzmann statistics and the Saha equation
(LTE case) or a collisional-radiative model (non-LTE case).  The energies
and pressures of the EOS tables are based on the free energy, thus
providing for thermodynamically consistent energy and pressure values.
For the non-LTE case,
the effects of photoionization and photoexcitation are ignored when computing
the temperature- and density- dependent tabular EOS data, as the radiation
field is a non-local quantity, {\em i.e.}, the radiation can originate in other
parts of the plasma, and therefore is unknown when generating the tables.  Detailed
atomic models are utilized in which all ionization stages are included, {\em e.g.},
for Ar, a total of $>10^4$ energy levels are included.
More than $10^6$ transitions were used for the opacity calculations.
Frequency-dependent radiation 
transport is treated using a radiation diffusion model; fifty frequency groups
were used in this work.  When modeling a plasma
that is not optically thick (our case)
with radiation diffusion, it is possible that radiation
losses could be overestimated.  Finally, we checked energy conservation
at $t=5$~$\mu$s (near peak compression)
for the polytropic, LTE tabular, and non-LTE tabular cases, using
run~6 of Table~\ref{eos-table} as an example.  Energy is conserved to
within 0.1\%, 0.1\%, and 1.1\% for the polytropic, LTE, and non-LTE runs,
respectively.  These are very small effects compared to the differences
we report in this paper, and thus we conclude that energy conservation does
not impact our results.

We previously verified\cite{awe11} a \helios\ calculation
against the convergent shock Noh problem\cite{noh87} and
got better than 5\% agreement for the highest resolution case we ran
(50~$\mu$m zone resolution over a sphere with 25~cm initial radius).
We also did a grid resolution convergence study showing
that an average grid resolution of 250~$\mu$m gave 
$\sim 10$\% accuracy.\cite{awe11}  In this work, we used 400 computational zones
over the initially 255~mm thick liner (average of 637.5~$\mu$m/zone), which was chosen
based on achieving a
balance between getting reasonable agreement in the benchmarking studies described
in Sec.~\ref{benchmarking} and the run time of each simulation (typically a few hours to
a day per run).  We used the automatic zoning feature of \helios, and thus the zone size
was much smaller than the average near boundaries and much larger elsewhere.

When using a polytropic EOS, \helios\ only allows use of a one-temperature (1T)
model.  The tabular EOS simulations allow for separate ion $T_i$ and electron
$T_e$ temperatures.    All reported \helios\
results with polytropic EOS are 1T, and all reported results with tabular EOS are 2T with
radiation diffusion.  To verify that the use of 1T versus 2T does not have a great
impact on our results, we compared the time evolution of $T_e$ and $T_i$ for
the zone initially at $R=25.1$~cm and saw no difference.
Furthermore, the $T_i$ evolution from a 1T run was virtually
the same as that from the 2T run.

\subsection{Benchmarking against results of Awe {\em et al.}}
\label{benchmarking}

Because Awe {\em et al.}\cite{awe11} used the 1D Lagrangian radiation-hydrodynamics
code \raven,\cite{oliphant81} we first benchmarked \helios\ against \raven\ by
performing runs 1--8 of Table~\ref{eos-table} using a polytropic EOS and
comparing with the \raven\ results.
In this paper, all polytropic EOS runs used $\gamma = 1.6667$.
We also included thermal tranport and radiation diffusion to be as consistent
as possible with Awe {\em et al.},\cite{awe11} which found that in
these spherically convergent simulations, non-physical temperature
and pressure spikes occur near the origin due to compression of the leading edge
material of the liner.  Including
thermal and radiation transport in the simulations helps prevent these non-physical
spikes, which prevent the imploding plasma liner from reaching as
small a radius as it otherwise would, leading to a much-reduced peak pressure.\cite{awe11}

\begin{figure}[!h]
\includegraphics[width=3truein]{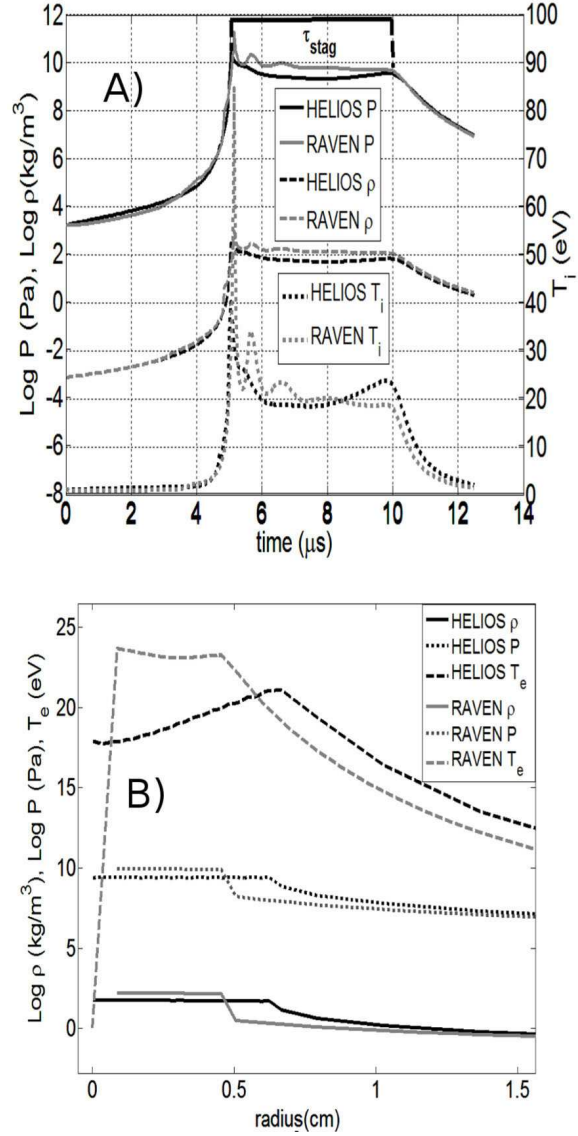}
\caption{Benchmarking comparisons of (A)~pressure $P$, mass density $\rho$, and ion
temperature $T_i$ 
(in the computational zone initially 1.0~cm from the inner edge of the liner)
versus time; and (B)~$P$, $\rho$, and electron temperature
$T_e$ versus radius at $t=6.5$~$\mu$s
between \raven\ and \helios\ argon polytropic EOS simulation results of 
run~6 of Table~\ref{eos-table}\@.}
\label{helios-raven-comparison}
\end{figure}

We compared the quantities $P_{stag}$ and $\tau_{stag}$,
where $P_{stag}$ is defined as the thermal pressure in the simulation
zone that is initially 1.0~mm from the inner edge of the
liner ($R=24.2$~cm) averaged over a time duration $\tau_{stag}$, which is defined 
(and also shown graphically in Fig.~\ref{helios-raven-comparison}A) as the
time duration between maximum pressure
to the time when pressure sustainment ends due to the outward going shock reaching
the outer edge of the incoming liner.
We found that \raven\ $P_{stag}$ results (Table~\ref{raven-table})
are 1.2--3.5 times higher than the
\helios\ results (Table~\ref{polytropic-table})\@.  
The $\tau_{stag}$ results agree to within 20\%.
These differences were considered to be small enough to proceed with this study.

We also compared \helios\ and \raven\ simulation results 
for the time evolution (Fig.~\ref{helios-raven-comparison}A) of 
the pressure $P$, mass density $\rho$, and ion temperature $T_i$ 
of a single computational zone (initially
1.0~cm from the inner edge of the liner, to be consistent with Fig.~4 of Awe {\em et
al.}\cite{awe11}), and the radial profiles (Fig.~\ref{helios-raven-comparison}B)
of $P$, $\rho$, and electron temperature $T_e$ at a single time ($t=6.5$~$\mu$s),
for run 6 of Table~\ref{eos-table}.
Both comparisons indicate good qualitative agreement with some discrepancies.
One difference is the increase in both $P$ and $T_i$
near the end of {\em $\tau_{stag}$} for the \helios\ but not the \raven\
results.  The magnitude of the rise in $P$ increases with increasing
$v_0$ and is present throughout the entire stagnated liner,
and is thus not a localized numerical effect in a single zone.  The
reason for this rise is not yet understood, but because
this does not affect the conclusions of this paper, we have
set the issue aside.  There are also differences in the radial profiles 
at $t=6.5$~$\mu$s,
largely due to the slight difference in time evolution between the two simulations.
At this time,
there is an outward going shock for both the \helios\ (at $\sim 0.6$~cm) and \raven\ 
(at $\sim 0.5$~cm) simulations, but
the \helios\ shock has propagated farther outward.  The discrepancy in $T_e$ inside
the shock also appears to be due to the difference in time evolution, as slightly later
radial
profiles from the \raven\ results also fall toward the origin (see Fig.~4 of
Awe {\em et al.}\cite{awe11}).
These differences do not affect the conclusions of this paper.
The rest of the paper focuses on comparing \helios\ simulations using 
a polytropic EOS versus tabular EOS data.

\section{Results}
\label{results}

This section presents \helios\ simulation results and analysis.
Firstly, we discuss LTE versus non-LTE tabular EOS results, showing that there are
only minor differences, thus justifying our choice to focus on the use of
LTE for the subsequent results in the paper.  Secondly, we compare argon
plasma liner implosions between polytropic and a LTE tabular EOS to
assess the effects of ionization and a more sophisticated treatment of EOS\@.
Finally, we explore the effects of using different plasma liner species for
run~6 of Table~\ref{eos-table}, again using LTE tabular EOS data.

\subsection{LTE versus Non-LTE EOS tabular data}

\begin{figure}[!h]
\includegraphics[width=3truein]{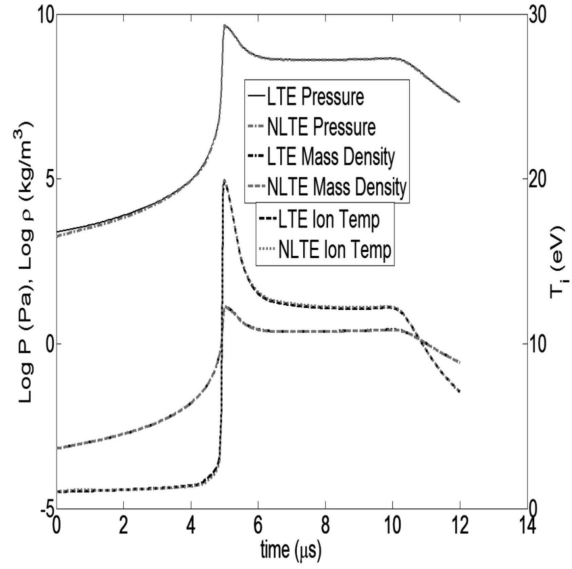}
\caption{Comparison of pressure $P$, mass density $\rho$, and ion temperature
$T_i$ (in the computational zone that is initially 1.0~cm inside the inner
edge of the liner) versus time between LTE 
and non-LTE argon simulation results for run~6 of Table~\ref{eos-table}.}
\label{lte-vs-nlte}
\end{figure}

The initial conditions we used (see Table~\ref{eos-table}) are in a parameter space
in which LTE is not obviously correct.  Non-LTE can have an effect
on EOS ({\em i.e.}, ionization/excitation) and radiation losses
(excited states tend to be less populated).  Below an argon ion density of
$\sim 10^{17}$~cm$^{-3}$ at $\sim 1$~eV, the fractions of
Ar~I, Ar~II, and Ar~III as a function of temperature are noticeably
different between LTE and non-LTE calculations. 
Above $\sim 10^{17}$~cm$^{-3}$,
corresponding to later stages of liner implosion and stagnation,
the EOS data are similar.
We performed run~6 of Table~\ref{eos-table} 
using both LTE and
non-LTE tabular EOS data to explicitly assess the difference in results.
Figure~\ref{lte-vs-nlte} shows a comparison
of $P$, $\rho$, and $T_i$ versus time for the
computational zone that is initially 1.0~cm from the inside edge of the
plasma liner.  Note that for the tabular EOS
runs, the pressure is calculated as $n_i T_i + n_e T_e$,
where $n_i$ and $n_e$ are the ion and electron densities, respectively,
$n_e=Zn_i$, and $Z$ is the mean charge state.
The LTE and non-LTE results are virtually indistinguishable
over the entire duration of the simulation, and there
are also no appreciable differences in $P$, $\rho$, and $T$ over
the entire liner at several time steps (not shown).
This indicates that either LTE or non-LTE
EOS models could be used without altering the relevant simulation results for
the purposes of this study,
and thus we used LTE for all subsequent results presented in the paper.

\subsection{Tabular LTE EOS versus polytropic EOS}
\label{EOS-vs-poly}

\begin{figure} [!h]
\includegraphics[width=3truein]{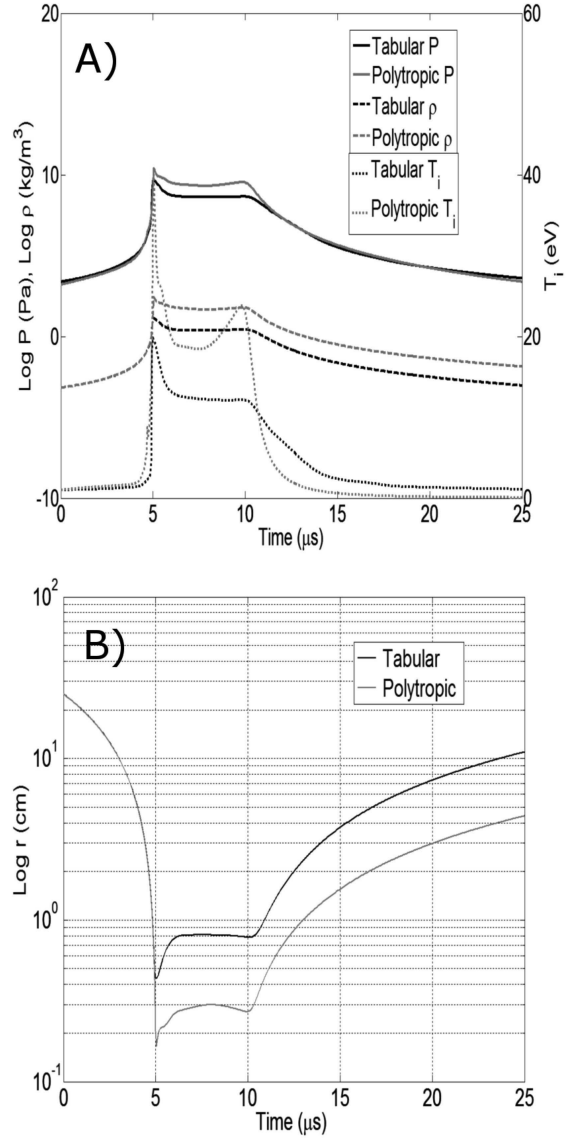}
\caption{Comparison between argon LTE tabular EOS and polytropic EOS simulation results
for run~6 of Table~I for the
time evolution of (A)~pressure $P$ and ion temperature $T_i$, and (B)~radial
position of the liner, for the computational zone that is 1.0~cm from the inside edge
of the liner.}
\label{eos-vs-polytropic}
\end{figure}

To assess the effects of ionization and other effects included
in the LTE tabular EOS data, we compared \helios\ simulation
results using LTE tabular EOS with those using a polytropic ($\gamma=1.6667$)
EOS for runs 1--8 of Table~\ref{eos-table}\@.
For the computational zone that is initially 1.0~mm from the inside
edge of the liner, the maximum pressure $P_{max}$ is 3.6--6.1 times
larger and $P_{stag}$ 3.9--8.6 times larger for the
polytropic EOS results.  Complete results for both the tabular LTE
and polytropic EOS's are shown in
Tables~\ref{eos-table} and \ref{polytropic-table}, respectively.
The $T_i$ and $\rho$ at peak compression were also higher in
the polytropic EOS results as well (not shown).
Figure~\ref{eos-vs-polytropic} shows $P$, $\rho$, $T_i$, and radial position
versus time of the computational zone initially at 25.1~cm (1.0~cm from the
inside edge of the liner) for the initial conditions of run~6 of Table~\ref{eos-table}\@.
The lower $P_{max}$ ($\sim$ zone thermal energy divided by zone volume)
for the tabular EOS result is consistent with the combined effects of lower zone
thermal energy ($\sim \rho T$) and higher zone volume at peak compression,
as compared to the polytropic result.
Next, we investigate in more detail the reasons for the
lower zone thermal energy and higher zone volume at peak compression for the tabular EOS
run.

\begin{figure}[!h]
\includegraphics[width=3truein]{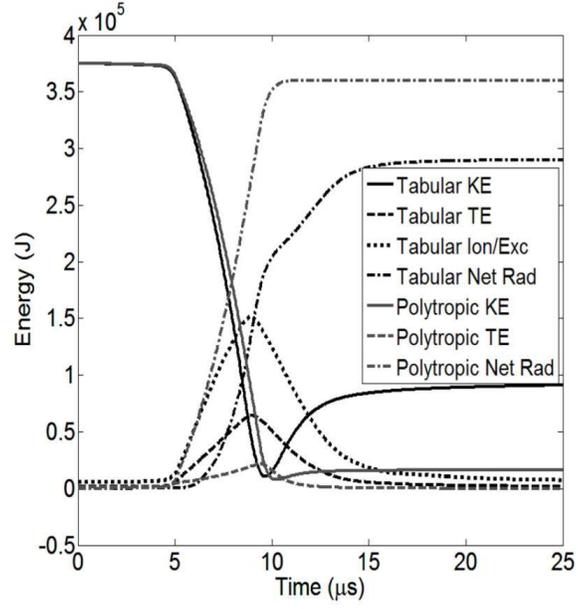}
\caption{Comparisons of the instantaneous total liner kinetic (KE), thermal (TE), 
ionization/excitation, and time-integrated
radiated energies of run 6 of Table~\ref{eos-table}
using argon tabular LTE and polytropic EOS models.  The latter does not include
ionization/excitation.}
\label{energy-vs-time-run-6}
\end{figure}

Figure~\ref{energy-vs-time-run-6} shows various components of the total
liner energy versus time for run~6 of Table~\ref{eos-table}\@.
The liner kinetic
energy $KE$ falls quickly around 5~$\mu$s when the leading edge of the
liner reaches peak compression and stagnates.  
In the polytropic EOS case, the 
difference between the initial $KE$ and the stagnated liner thermal energy $TE$ is
due mostly to radiative losses.  In the tabular EOS case,
around half of the initial liner $KE$ goes to ionization/excitation
energy, which is stored in free and excited electrons in the stagnated 
liner.
The mean charge peaks around 7
(see Sec.~\ref{species}).
The result of including the ionization/excitation physics
is a drastic reduction in the liner compression at stagnation, leading
to lower $P_{stag}$ (see Tables~\ref{eos-table} and \ref{polytropic-table}).

\subsection{Stagnation pressure scalings}

Awe {\em et al.}\cite{awe11} found the approximate scalings $P_{stag}\sim v_0^{15/4}$ and
$P_{stag}\sim n_0^{1/2}$ for
the cases in Table~\ref{eos-table} using a polytropic EOS model,
and gave an heuristic argument for why $P_{stag}\sim v_0^5$ ought to
be an upper bound on the scaling dependence of $P_{stag}$ on $v_0$.
The heuristic argument invoked the following
assumptions:  (1)~perfect conversion of
liner $KE$ to liner $TE$ at stagnation,
(2)~no radiation and thermal losses, and (3)~no entropy production
due to the outward going post-stagnation shock.
Because the \raven\ runs of Awe {\em et al.}\cite{awe11} included radiation and thermal
losses, the observed weaker $P_{stag}\sim v_0^{15/4}$ scaling is consistent
with the heuristic upper bound of $P_{stag}\sim v_0^5$.
Our \helios\ runs using a LTE tabular EOS violate assumption (1) above due
to the major ionization/excitation energy channel discussed in 
Sec.~\ref{EOS-vs-poly}\@.  Thus, it is expected we should see an even
weaker scaling than $P_{stag}\sim v_0^{15/4}$.

\begin{figure}[!h]
\includegraphics[width=3truein]{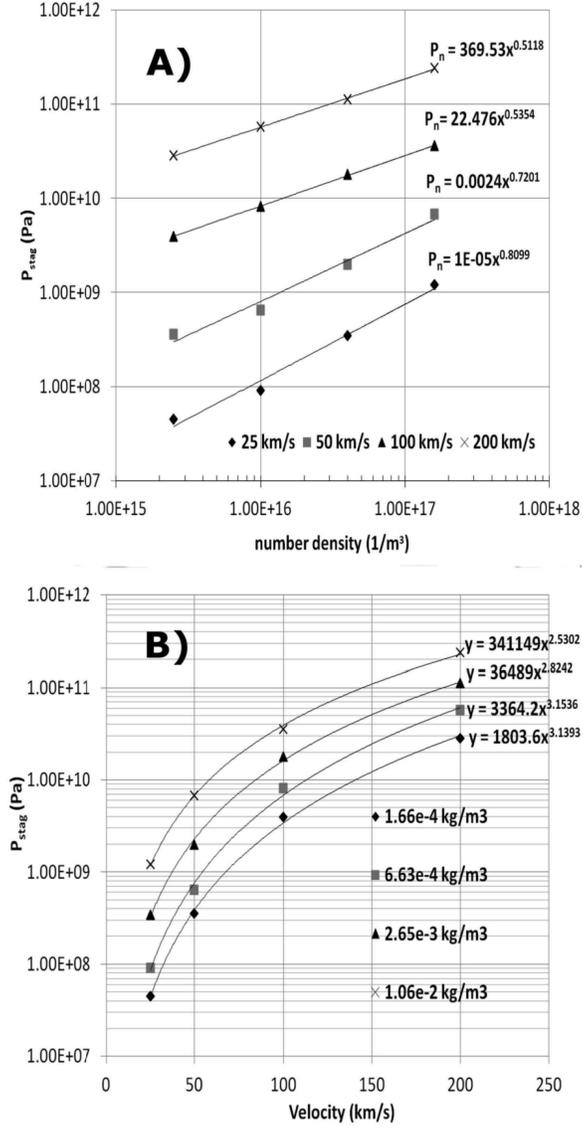}
\caption{$P_{stag}$ versus (A)~$n_o$ and (B)~$v_o$ for the argon runs in
Table~\ref{eos-table}, and corresponding power law fits.}
\label{scalings}
\end{figure}

Figure~\ref{scalings} shows $P_{stag}$ versus
$n_0$ and $v_0$ and fits to power law functions for \helios\ simulations
of the cases in Table~I using a LTE tabular EOS model.
While the fitted exponents are very similar across
density and velocity groups in Awe {\em et al.},\cite{awe11} {\em i.e.},
$P_{stag}\sim n_0^{0.54\pm 0.02}$ and $P_{stag}\sim v_0^{3.71 \pm 0.08}$,
our fitted exponents have a much wider spread, {\em i.e.},
$P_{stag}\sim n_0^{0.64 \pm 0.14}$ and $P_{stag}\sim v_0^{2.91 \pm 0.30}$.
This larger spread in exponents is likely due to the fact that 
the LTE tabular EOS is sensitive to density and temperature in the imploding
plasma liner, thus introducing a degree of variability that depends on the liner
initial conditions not expected
in the polytropic EOS simulations of Awe {\em et al.}\cite{awe11}
Our $P_{stag} \sim v_0^{2.91 \pm 0.30}$ scaling (depending on $n_0$) is indeed weaker than 
the approximate $P_{stag}\sim v_0^{15/4}$ scaling of Awe {\em et al.},\cite{awe11}
as expected.
Our $P_{stag}\sim n_0^{0.64 \pm 0.14}$ scaling agrees with the $P_{stag}\sim n_0^{1/2}$
scaling of Awe {\em et al.}\cite{awe11}
within our observed variation across velocity groups.

\subsection{Varying the liner species}
\label{species}

Due to the significant effect of ionization/excitation on
achievable $P_{stag}$, we explored the use
of different plasma species
for the same initial liner $\rho_0$ and $v_0$ (and hence $KE$)
of run~6 of Table~\ref{eos-table}\@.
The only quantity that was varied
was the initial number density $n_0$ to keep $\rho_0$ constant.
We performed \helios\ simulations with LTE tabular EOS data to investigate
the effect of species on $P_{stag}$.
The species chosen for study are those of interest 
as fusion fuel, blanket materials, or MIF implosion drivers for fusion
energy applications:
H, D-T (50\%-50\% by atom), $^4$He, $^6$Li, $^{11}$B, Ne, Ar, Kr, and Xe.

\begin{figure}[!h]
\includegraphics[width=3truein]{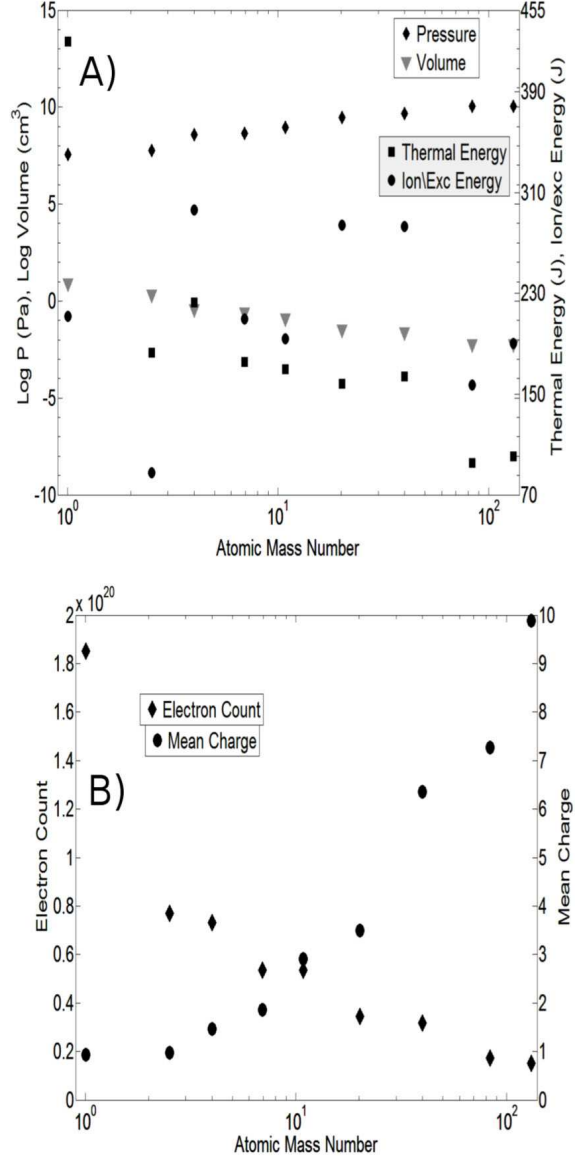}
\caption{(A)~Pressure, volume, thermal
energy, and ionization/excitation energy, 
and (B) total electron count and mean charge versus liner atomic mass number at the
time of peak compression,
for the computational
zone initially 1.0~cm from the inner edge of the liner, and for initial liner
$v_0$ and $\rho_0$
corresponding to run~6 of Table~\ref{eos-table}\@.
Liner species
(from left to right on plot) include H, D-T, $^4$He, $^6$Li, $^{11}$B, Ne, Ar, Kr,
and Xe\@.}
\label{species-vs-amu}
\end{figure}

Figure~\ref{species-vs-amu}A shows \helios\ simulation results
using a LTE tabular EOS model for $P$, volume,
$TE$, and ionization/excitation energy at
peak compression versus
atomic mass number for the zone initially 1.0~cm from the inner edge of the liner.
There is a clear trend of increasing peak $P$ with atomic mass number,
quantitatively consistent with the combined effects of decreasing volume and thermal
pressure with atomic mass number at peak compression.  The thermal and
ionization/excitation energies are not monotonic with atomic mass number, and this
is related to the different atomic and radiative properties of the different species.

We also examined the mean charge and
electron count versus time (Fig.~\ref{species-vs-amu}B)
for the same computational zone as above.
Although the
heavier species have more electrons to ionize, there are a larger
number of the lighter species present by the ratio of their
atomic mass numbers.  It turns out that the peak mean 
charge for the heavier species reaches only about 5--10 (Fig.~\ref{species-vs-amu}B),
much less than the mass ratios between the heavier and lighter species,
and thus the electron count is higher for the lighter species.  Thus, the
ionization/excitation energies are in general higher for the lighter species, in part
explaining why they reach lower peak pressures for the initial conditions imposed.

\section{Discussion and summary}
\label{summary}

It appears that the key effect of using a LTE tabular EOS rather than a polytropic EOS is
the significant diversion of liner $KE$
into ionization/excitation, which is not included in the polytropic EOS model.
The diversion of initial liner $KE$ leads to less effective compression,
larger minimum volume at peak compression, and thus lower $P_{stag}$ which
is inversely proportional to volume.
This is detrimental for reaching very high peak temperatures and
pressures in targetless spherical plasma liner implosions, which
was one of the design objectives of PLX\@.  For
a given initial liner $KE$, which is a natural experimental constraint,
the results suggest the use of lower density and higher velocity to achieve
higher peak pressure.  The previous statement
is confirmed by examining, {\em e.g.}, the $P_{stag}$ results from
runs 3, 6, and 9 (375~kJ), or runs 4, 7, 10, and 13 (1.5~MJ) of
Table~\ref{eos-table}\@.  Higher
velocity should also be coupled with the use of lower atomic mass number 
species so as to fully strip the ions and eliminate the ionization/excitation
and line radiation energy channels beyond a certain point in the implosion,
after which the liner $KE$ will predominantly go toward stagnation thermal
pressure.
This is in apparent contradiction with the results of
Sec.~\ref{species}, which suggest the use of
heavier species for higher $P_{stag}$ but
only because we constrained the species comparison study by
enforcing the same $v_0$ and $\rho_0$.  In our species comparison study,
the much higher $n_0$ of the lighter species leads to
more of the initial liner $KE$ going into ionization/excitation.
Further study is needed for optimizing $P_{stag}$ in targetless spherical
plasma liner implosions for a given initial liner $KE$.

The above discussion does not apply for the MIF application
of a plasma liner compressing a magnetized D-T target plasma.\cite{hsu12}
In that case, the imploding plasma liner is intended to be only a pusher that
should complete its job of compressing the D-T target to fusion conditions
before the liner itself has a chance to get significantly compressed and heated.
Thus, for initial liner $n_0$ and $v_0$
that are likely to be limited by available plasma gun technology, the MIF
application likely will prefer a heavy liner species
(to maximize $\rho v^2$) since the ionization/excitation
should not matter until after the D-T target has reached peak compression.

In summary, we have presented new 1D \helios\ 
radiation-hydrodynamic simulation results of
spherical plasma liners imploding on vacuum using a LTE tabular EOS model.
The results were compared in detail to \helios\ simulations using
a polytropic EOS model ($\gamma=1.6667$), which in turn were benchmarked
against the polytropic EOS simulation results
of Awe {\em et al.}\cite{awe11}  We found that
the liner stagnation pressure $P_{stag}$
is lower by a factor of 3.9--8.6 using the LTE tabular EOS model.  This is attributed
to a significant amount of the initial liner kinetic energy going toward
ionization/excitation of the plasma liner, resulting in less effective compression
of the liner.  We also studied the effect of different species on $P_{stag}$
for the same initial liner geometry, mass density, and velocity.  We found
that with these constraints,
the heaviest species reach the highest values of $P_{stag}$.  This is
attributed to less total energy going to ionization/excitation due to a smaller
value at peak compression of the liner mean charge times the initial number density.

\begin{acknowledgments} The authors thank T. J. Awe for access to the
\raven\ simulation data.  This work was supported by the Office of Fusion Energy 
Sciences of the U.S. Department of Energy.  
\end{acknowledgments}



%


\newpage

\begin{table}[!h]
\begin{tabular}{ccccccccccc}
\hline
\hline
Run & $n_0$ (cm$^{-3}$) & $\rho_o$ (kg/m$^3$) & $v_0$ (km/s) & $KE_0$ (J) & 
$\tau_{stag}$ ($\mu$s) & $P_{stag}$ (Pa) & $P_{max}$ (Pa) & $P_{stag}\tau_{stag}$ & 
$P_{stag}\tau_{stag}/KE_0$\\
\hline

1 & 2.5E+15 & 1.66E-4 & 25 & 2.35E+4 & 9.40 & 4.49E+07 & 1.31E+08 & 4.22E+2 & 0.018\\
2 & 2.5E+15 & 1.66E-4 & 50 & 9.39E+4 & 5.35 & 3.54E+08 & 1.84E+09 & 1.90E+3 & 0.020 \\
3 & 2.5E+15 & 1.66E-4 & 100 & 3.76E+5 & 2.70 & 3.93E+09 & 1.43E+10 & 1.06E+4 & 0.028 \\
4 & 2.5E+15 & 1.66E-4 & 200 & 1.50E+6 & 1.30 & 2.84E+10 & 5.69E+10 & 3.69E+4 & 0.025 \\
5 & 1.0E+16 & 6.63E-4 & 25 & 9.38E+4 & 9.55 & 9.15E+07 & 3.57E+08 & 8.74E+2 & 0.093 \\
6 & 1.0E+16 & 6.63E-4 & 50 & 3.75E+5 & 5.35 & 6.43E+08 & 4.63E+09 & 3.44E+3 & 0.092 \\
7 & 1.0E+16 & 6.63E-4 & 100 & 1.50E+6 & 2.70 & 8.20E+09 & 3.13E+10 & 2.21E+4 & 0.015 \\
8 & 1.0E+16 & 6.63E-4 & 200 & 6.00E+6 & 1.35 & 5.70E+10 & 1.26E+11 & 7.40E+4 & 0.013 \\
9 & 4.0E+16 & 2.65E-3 & 25 & 3.75E+5 & 10.45 & 3.47E+08 & 1.43E+09 & 3.62E+3 & 0.010 \\
10 & 4.0E+16 & 2.65E-3 & 50 & 1.50E+6 & 5.45 & 2.00E+09 & 1.43E+10 & 1.09E+4 & 0.007 \\
11 & 4.0E+16 & 2.65E-3 & 100 & 6.00E+6 & 2.70 & 1.79E+10 & 1.30E+11 & 4.83E+4 & 0.008 \\
12 & 4.0E+16 & 2.65E-3 & 200 & 2.40E+7 & 1.35 & 1.13E+11 & 7.39E+11 & 1.53E+5 & 0.006 \\
13 & 1.6E+17 & 1.06E-2 & 25 & 1.50E+6 & 10.50 & 1.23E+09 & 4.97E+09 & 1.29E+4 & 0.009 \\
14 & 1.6E+17 & 1.06E-2 & 50 & 6.00E+6 & 5.70 & 6.81E+09 & 6.02E+10 & 3.88E+4 & 0.006 \\
15 & 1.6E+17 & 1.06E-2 & 100 & 2.40E+7 & 2.70 & 3.56E+10 & 4.27E+11 & 9.61E+4 & 0.004\\
16 & 1.6E+17 & 1.06E-2 & 200 & 9.59E+7 & 1.35 & 2.40E+11 & 2.23E+12 & 3.24E+5 & 0.003\\
\hline
\hline
\end{tabular}
\caption{Initial conditions (same as Table~\ref{eos-table}I of Awe {\em et al.}\cite{awe11}) and 
summary of \helios\ argon simulation results for the indicated quantities,
using LTE tabular EOS data.  The results are for
the computational zone that is initially 1.0~mm inside the inner edge of the
liner.  All cases have initial liner inner radius of 24.1~cm,
liner thickness of 25.5~cm, and temperature of 1.0~eV\@.}
\label{eos-table}
\end{table}

\newpage

\begin{table}[!h]
\begin{tabular}{ccccccccccc}
\hline
\hline
Run & $\tau_{stag}$ ($\mu$s) & $P_{stag}$ (Pa) & $P_{max}$ (Pa) & $P_{stag}\tau_{stag}$ &
$P_{stag}\tau_{stag}/KE_0$\\
\hline
1 & 8.82 & 2.37E+08 & 3.17E+9 & 2.09E+3 & 0.09 \\
2 & 4.51 & 3.11E+09 & 8.58E+10 & 1.40E+4 & 0.15 \\
3 & 2.29 & 3.43E+10 & 8.82E+11 & 7.84E+4 & 0.21 \\
4 & 1.03 & 6.29E+11 & 4.26E+13 & 6.48E+5 & 0.43 \\
5 & 8.84 & 4.78E+08 & 4.41E+09 & 4.23E+3 & 0.05 \\
6 & 4.56 & 6.79E+09 & 1.32E+11 & 3.09E+4 & 0.08 \\
7 & 2.33 & 8.39E+10 & 2.40E+12 & 1.95E+5 & 0.13 \\
8 & 1.19 & 8.32E+11 & 2.35E+13 & 9.86E+5 & 0.16 \\
\hline
\hline
\end{tabular}  
\caption{\raven\ argon simulation results using a polytropic EOS ($\gamma=1.6667$)
for runs 1--8 of Table~\ref{eos-table} (reproduced from 
Table~II of Awe {\em et al.}\cite{awe11}).
The results are for
the computational zone that is initially 1.0~mm inside the inner edge of the
liner.}
\label{raven-table}
\end{table}

\begin{table}[!h]
\begin{tabular}{ccccccccccc}
\hline
\hline
Run & $\tau_{stag}$ ($\mu$s) & $P_{stag}$ (Pa) & $P_{max}$ (Pa) & $P_{stag}\tau_{stag}$ &
$P_{stag}\tau_{stag}/KE_0$\\
\hline
1 & 10.30 & 1.73E+08 & 7.44E+08 & 1.79E+3 & 0.08 \\
2 &  5.05 & 1.87E+09 & 1.09E+10 & 9.44E+3 & 0.10 \\
3 &  2.55 & 1.63E+10 & 5.15E+10 & 4.16E+4 & 0.11 \\
4 &  1.30 & 1.78E+11 & 3.19E+11 & 2.32E+5 & 0.15 \\
5 & 11.15 & 3.91E+08 & 1.64E+09 & 4.36E+3 & 0.05 \\
6 & 5.05 & 4.11E+09 & 2.71E+10 & 2.08E+4 & 0.06 \\
7 & 2.65 & 7.09E+10 & 1.54E+11 & 1.88E+5 & 0.13 \\
8 & 1.30 & 3.86E+11 & 7.70E+11 & 5.02E+5 & 0.08 \\
\hline
\hline
\end{tabular}  
\caption{\helios\ argon simulation results using a polytropic EOS ($\gamma=1.6667$)
for runs 1--8 of Table~\ref{eos-table}, for
benchmarking against the results of Awe {\em et al.}\cite{awe11}
The results are for
the computational zone that is initially 1.0~mm inside the inner edge of the
liner.}
\label{polytropic-table}
\end{table}

\newpage

\end{document}